# Evaluating User Experience in Literary and Film Geography-based Apps with a Cartographical User-Centered Design Lens

Min Rezaei[1]*

PhD candidate, College of Agricultural and Environmental Sciences, University of California, Davis

minrezaei@ucdavis.edu

Patsy Eubanks Owens

Professor, College of Agricultural and Environmental Sciences, University of California, Davis

peowens@ucdavis.edu

Darnel Degand

Professor, School of Education, University of California, Davis

ddegand@ucdavis.edu

Geography scholarship currently includes interdisciplinary approaches and theories and reflects shifts in research methodologies. Since the spatial turn in geographical thought and the emergence of geo-web technologies, geography scholarship has leaned more toward interdisciplinarity. In recent years geographical research methods have relied on various disciplines ranging from data science to arts and design. Literary geography and film geography are two subfields of geography that employ novels and films in exploring spatiality, respectively. In addition to geographical concepts, these courses include many aspects of relations in space, including human-human relations, human-environment relations, et cetera, which were barely addressed in traditional geography courses. However, a review of the employment of geo-web technologies in literary and film geography practices reveals that these practices have mostly remained limited to isolating "geographical" passages from novels or movies. This paper explores new opportunities for designing film and literary geography-based apps using a cartographical user-centered design framework.

## 1. INTRODUCTION

In the twentieth century, with the rejection of the values of the top-down approach of modernism by geographers, new lines of study emerged which relied less on scientific measurements and more on philosophy, poetics, and sociology in exploring humans and their spaces. In the 1960s, with the rise of theories of postcolonialism, Edward Soja [1] made a reassertion to the concept of space by using "spatial turn" in humanities and other fields related to the built environment such as architecture and urban design. The spatial turn in the humanities indicates that space is neither a subjective nor objective experience. It maintains that social and cultural dynamics of power relations shape the space [2,1,3,4]. Proponents of the spatial turn assert that the traditional map is a form of institutionalization that misses feelings, memories, stories, et cetera [5,6,7,8,9,10]. They believe that lines and points cannot represent the actual act of walking, window shopping, or wandering. Exploring this "lived experience" of people made geographers more interested in employing materials such as diaries, memories, travelers' notes, and novels. In recent years, geographic information technologies have also contributed profoundly to understanding the "spatial turn" in the humanities. These technologies provide the opportunity to create "spatial stories" of people's lived experiences [9,10,11,12,8].

Partly because of the emergence of spatial turn theory in geography, and partly due to the emergence of geo-web technologies, geography scholarship has shifted toward more interdisciplinary studies in recent years. Also, some argue that university-based geography programs have experienced a decline in enrollment mainly because geography-related jargon is outdated and being severely outcompeted by environment- and sustainability-related language [13]. Since the spatial turn, geographical research methods have relied on a plethora of disciplines, ranging from data science to arts and design [10]. Literary geography and film geography are two subfields of geography which rest their epistemology on using novels and films in exploring spatiality, respectively. These two fields have been well-established in academia in different parts of the world. In addition to geographical concepts, these courses include many aspects of relations in space, including man-man relations, man-environment relations, et cetera, which were barely addressed in traditional geography courses. However, a review of employing geo-web technologies in literary and film geography practices reveals that these practices have mostly remained limited to isolating "geographical" passages from novels or movies. This paper explores new opportunities for designing film and literary geography-based apps using a cartographical user-centered design framework.

### 1.1. The Spatial Turn and Geo-Web Technologies

Before postmodernism, dominant thought in the geographical realm considered maps as "scientific'' tools representing power [14,15]. As Harley [15:282-283] says, "as much as guns and warships, maps have been the means of imperialism." These graphic representations of boundaries and territories legitimized the victories and conquests of rulers as well as ownerships of individuals and national properties, manifested ideologies of religious groups, and provided sensitive knowledge for the military [15,16]. While those static maps remain fruitful in geographical studies, they leave no room for user engagement in their creation or alteration. Even cartographers were not independent creators for the so-called traditional maps. They mainly served individuals, states, or market ambitions [15]. In the late twentieth century, with the rise of postmodern thought and the spatial turn, a new approach to the concept of space was introduced. Postmodernists believed that space is influenced by different power relations and is under constant change over time. Therefore, exploring these shifts can not be limited to employing observation and positivist approaches [17,18,19,20]. Postmodernists argue that traditional maps are a form of institutionalization that misses feelings, memories, etc. They [10,1,23,24] argue that traditional maps—which are mainly produced by engineers, planners, technocrats, and scientists—overlook the human side of space and downsize it to an object that can be represented by dots and other geometrical shapes [5,21,22]. These new lines of thought discern between "space" and "place" and demonstrate that space is a practiced place [5]. For instance, according to de Certeau [5], pedestrians transform the street from a place that is "geometrically defined by urban planning" into a space that is useful or pleasurable for them. Lefebvre [21], whose ideas had a huge influence on postmodern spatial studies, recognizes three different spaces to explain the complexities of exploring "space" — perceived space, conceived space, and lived space (see Figure 1).

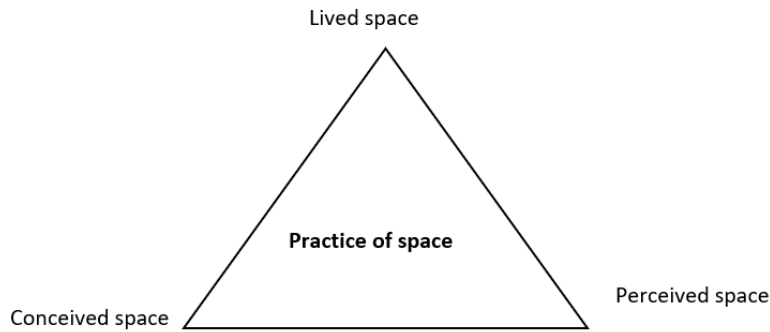

Figure 1: trilogy of space, based on Lefebvre's production of space [21]

According to Lefebvre [21], perceived space is the space of day-to-day lives, a space of daily reality and urban reality, and those realities shape people's mental image of space. Conceived space is tied to the spaces of power, such as spaces of technocrats and scientists, and includes maps, plans, rules, and regulations. Lived space or social space incorporates social actions [21]; it is the space of resistance against authorized power—the space of freedom—which users can manipulate without being controlled or watched. Postmodernists suggested that understanding these complexities requires borrowing methodologies from other disciplines such as the arts and humanities. They emphasized the human side of geography and became more inclined to employ non-scientific materials such as paintings, diaries, travelogues, memories, novels, and films to investigate the lived experience of people in geographical locations [25,21,1,82]. Reading space through narratives allows users to explore space in a broader way than linear chronological order [25,21]. Advocates of the "spatial turn" believe that people's lived experiences resonate more with narrative time [25,21,1,82]. In narrative time, one thing happens not just after something else (chronological order), but because something else occurred in a story or history [25]. Lefebvre's [21] "Production of Space'' was one of the earliest studies that used literary analysis to read urban spaces and places and proposed revolutionary methods in exploring space using literary texts. Lamme [26] recognizes four types of spatial knowledge that can be acquired while reading novels: first, landscape, which includes both physical and human landscapes; second, human ecology, which is about people-space interconnections; third, strategy, which is authorship over space by tools like maps, plans, et cetera; and, finally, regionalism, which means various geographical borders and realities.

Films have also been a popular medium in postmodern geography scholarship. Cresswell and Dixon [27] assert that: "While the very idea of landscape study is built around observation from a fixed point of a static scene, film viewing involves the observer taking a mobile view on a mobile world." Films can demonstrate both "real" (pieces of stuff outside the screen) and "reel" (pieces of stuff on the screen). Film geography was introduced as a subdiscipline of geography by Lukinbeal and Zimmermann (2006) [28]. They recognized four trajectories. First, geopolitics, which explains how film images and narratives are a part of a larger geopolitical imaginary; second, cultural politics, which is how socio-spatial meaning is displayed in the film; third, globalization, which situates the impact of economic imperatives of film production; and fourth, the trajectory of film geography research, science, and representation, which explores how films portray reality and how this reality is under the shadow of the dominant ideology and power.

Geographical representations in novels and films go beyond illustrating physical places. They encompass a wide range of "spatial" information such as human relations in spaces; their relation to the spaces; and their memories, fears, and everyday life practices in the space. Building upon interdisciplinary research trends, several geography programs have started to employ multi-faceted methods such as using films and novels in teaching spatial sciences to students in different majors all over the world [29-53].

Additionally, in the last two decades, the development of geo-web technologies and geographic information systems (GIS) revolutionized the mapmaking industry by adding interactivity to map creation. Using interactive maps helps blend oral stories, novels and stories, music, or other art forms to represent different representations of space [54]. It also provides the opportunity for cartographers to incorporate different layers of memories and other social, cultural, and historical aspects of space into the maps [8]. Interactivity challenged the top-down approach in producing maps and fostered the democratization of cartography by empowering users to create their own spatial representations and facilitating their participation in the planning and design of their communities [54,55,56,57,58,59].

In the late 1990s, as one of the consequences of technological development in cartography, users and usability were brought to the forefront of cartographical research [60]. In the 2000s, with the establishment of a new Commission on Use and User Issues of the International Cartographic Association (ICA), usability of map-based products became more important [61]. While there have been several studies on the usability of map-based apps and web maps, they mostly have focused on how previous map users make maps, interact, and explore with map apps [65,63]; User Interface design, which includes (map) representations of reality on a small display screen [64, 65]; geovisualization and users' interaction (map features) and mental maps [64]; and map display and graphical user interface design of mobile and web maps [61, 60]. Although there has been an emphasis on the involvement of cartographic expertise in cartographic design [63], research around the role of the cartographer and cartographical content in user-centered design for interactive mobile and web maps is limited.

## 2. METHODS

This research has been conducted to answer these two questions:
**What are the main themes that are included in literary or film geography course syllabi in academia?**
**Do apps related to literary geography and film geography reflect educational information around space**?
To answer the first question, twenty-five courses related to film and literary geography were reviewed. The courses were found by searching university course directories in geography, urban planning, urban studies, media and film studies, literature, and art history departments; the Google search engine, and the *Journal of Literary Geography.* The keywords we searched were "teaching film geography," "literary geography," "literature and the city," "film and the city," "cinema and the city," and "digital humanities." The search was expanded to English-speaking universities all over the world to create a geographically inclusive database. Twenty courses were found. The snowball method was used by consulting scholars in literary geography, film geography, and the digital humanities to discover if they teach courses related to this topic or knew of any related course that we had missed. Ten other syllabi were found. The criteria for choosing a syllabus was that it had to have assignments, learning objectives, and course content sections. Five syllabi were eliminated as they did not meet the requirements. Then the syllabi were analyzed using the MAXQDA 2020 code system. Each syllabus was read manually and the content including the references, schedule, and introduction was coded based on the three main categories of space by Henri Lefebvre. Subcategories which were defined based on theories of the "spatial turn." The assignments and learning objectives were also coded based on authors' knowledge of spatial science and teaching experiences.

For the second question, to find apps related to literary geographies, Google Scholar, Elsevier, the ACM digital library database, a GIS history database, and the *Literary Geographies Journal* were used. The phrases which were used in the search were chosen to cover all the themes in the fields of literary and film geography. The keywords included "literary," "literature," "novel," "story," "film," "movie," and "deep mapping," combined with the word

"app" or its equivalent variants including "geospatial platform," "digital platform," "digital interface," et cetera. Five apps were found. By employing the snowball method, the sample was expanded to eighteen interactive web maps and apps. We decided to ignore 10 of the web maps because they had similar characteristics. Two apps were omitted because they could not maintain funding and failed to get into the market. Finally, after these exclusions, we had 6 digital platforms with unique characteristics.

A semi-structured 1:1 interview with the app and interactive map developers through Zoom was conducted and a questionnaire was sent via email to those who were not available to meet on Zoom. Additionally, articles, videos, and website reports related to the apps or interactive maps were reviewed. The app contents were then analyzed based on: 1) team expertise — to see how they used different experience backgrounds and how it leads to a more user-centered experience for the digital platform, 2) aim — to explore why they designed the app, 3) user interactivity — to explore how the users can interface with the map and what content they will be exposed to, and 4) methods of data collection of each app or web map — to understand the extent of the research behind the app or interactive map. Based on the analysis, a spatial user centered design (SUCD) framework for designing literary and film-based maps with a cartographical lens was proposed. The framework links literary and film geographic theories to digital practices in order to help people explore "space" with all of its complexities. Finally, a model of a literary/film app based on the SUCD model was proposed.

**3. Analyzing Literary and Film Geography Course Syllabi**

Twenty-five syllabi in three different groups including "literature and city," "cinema and city," and "literature and film" were imported to the MAXQDA 2020 software. The software allows researchers to do hermeneutical analysis by coding different segments of the text. It also helps researchers to understand the relations between the defined codes [66]. The three main code categories, as mentioned, were perceived space, conceived space, and lived space. After close reading the courses, the main categories were broken down into different subcategories, as shown in Tables 1-

Table 1- codes for Conceived Space

| Conceived space | |
|---|---|
| Codes | Segments |
| Media as Representational Tool | The first aim is to consider film as a geographic practice<br>How have filmmakers developed a visual language to evoke viewers' experiences of inhabiting and moving through space, to transport them to new places? [29]<br>Optimistic and pessimistic portrayals of cities and city living are presented in classic and modern films [30]<br>Cinema both emerged out of modern urban life and provided spectators with new ways to see and reflect on urban experience<br>This class investigates the interaction between cinema and the city by probing the way that film imaginatively engaged with urbanity as both an evolving physical and technological environment and a way of life<br>How urban representation has been sustained and transformed within particular international film genres and modes, including the city symphony, slapstick, noir and detective thrillers, the musical, and science fiction [31]<br>The twentieth century's ideas of the city are closely tied to its representation in the paradigmatic modernist medium of film, and the course will illustrate changing understandings of urban space over time [32]<br>In return, the film industry has held a mirror up to the American city, sometimes enhancing, sometimes distorting, sometimes oversimplifying its complexities, exploring its realities, confirming and disconfirming its myths, always adding to the lore of urban life and influencing the popular consciousness of it<br>The beginning of film as a medium of mass entertainment emerged in the last years of the 19th century, contemporaneous with the maturation of many American cities into their current metropolitan form [33]<br>How are cities represented in films? [34]<br>Using film as a lens to explore and interpret various aspects of the urban experience in both the U.S. and abroad, this course presents a survey of important developments in urbanism from 1900 to the present day [35]<br>Students will explore the ideas and representation of the city and the urban experience in literary texts from diverse cultural contexts. [36]<br>We will probe the city by way of its literature and texts and examine the complementary reflection of a direct experience of the space found in the books [37]<br>It is a course on how cities are represented in cinema, and a variety of attitudes toward cities in different films will be explored<br>what attitude the film takes toward the presence of the city.<br>It is a film course, and will try to show how films create feelings and express meanings through the materials of cinema, compositions and camera movements and light, rhythm and editing and sound [38])<br>Throughout the course we will examine the ways in which New York City as a place, and as a place composed of a myriad of places, is represented through the camera's lens and the director's vision [39]<br>Representations of cities in art, literature, and film [40]<br>What does it mean to read or write a city? How might your answer change depending on whether the focus is on urban architecture, people? history, climate, design/planning, cultural production and institutions, political activity, lived everyday experience, etc.?<br>Our seminar seeks to explore their connection as it relates to the emergence and global spread of the modern and contemporary city<br>How has the spatial and social organization of the modern city informed the thematic and formal choices writers make?<br>How do the density and scale of the urban built environment impinge on the way writers view the world and tell their stories? What genres seem best suited to rendering urban life? Is the city the defining context of modern literature or its implicit if barely human hero? To what extent is the global diffusion of the novel form related to the growth of urbanization? [41]<br><br>The final strand will explore how and why the changing nature of the modern city and cityscape has shaped and variously altered our readings of texts (and understanding of textuality) and vice versa. [43] |

Table 1- codes for Conceived Space

| Codes | Segments |
|---|---|
| | Conceived space |
| Cinematic Techniques | We will closely consider the ways that various techniques of framing, camera movement, cutting, and assemblage of elements (mise-en-scène) construct cinematic space [44]<br>It is a film course, and will try to show how films create feelings and express meanings through the materials of cinema, compositions and camera movements and light, rhythm and editing and sound [38]<br>Modernism and Montage [31]<br>Such cinematic effects as montage, movement and subjective camerawork, combined with the organizing structures of narrative, will be seen as integral tactics toward understanding the city as a complex, lived environment [32]<br>How the filmmaking process (camera movements, lighting, dialogue, acting, etc.) is used as a method to describe space (filmmaking as a geographic method [39] |
| Skyscrapers | Views from Above and Below, Changing conceptions of the Modern City [45] |
| Suburbs | Suburbs Vs Cities [30]<br>Subcultural Space, Penal Space, Spatial Appropriations<br>American Derivé – The Suburbs and Alternative Spatial Practices, The Dualized City [43]<br>Long parallel lines of horsecar and streetcar racks pushed out from the city centers to the open land where residential suburbs began to grow<br>Myths and realities of the American suburb [33]<br>The urban conquest; the machine in the garden [33] |
| The Farm and the City | |
| Displacement and Immigration | Political Displacement via Gentrification (Emory University)<br>Narrative of Decay and Urban Blight (Urban Renewal in 1960s) [34]<br>Houselessness and gentrification [39]<br>Dispersal and Houslessnes [31]<br>Sizzling New York – Vibrancy and the Melting Pot<br>This course will investigate a wide range of works based in the city of New York, from the early days of the U.S. republic to the heyday of immigration and the terror of 9/11 [46]<br>Changing Demographics and Immigration [47]<br>Immigration and national identity<br>Shifting borders and global migration, and identities marked by legacies of slavery and colonization. We will also attend to the many ways in which contemporary city novels question the lines between local and global, home and exile. [42]<br>Industry moved in downtown, and the middle class moved out, leaving their own houses or properties to be occupied by foreigners and migrants from the countryside [33] |

Table 1- codes for Conceived Space

| Codes | Segments |
|---|---|
| | **Conceived Space** |
| Climate Change | Climate change and sustainability (Rutgers University)<br>Climate change and super storms<br>Environmental degradation<br>How distinct are the lines between urban/environmental and the built environment versus natural environments? How does climate change challenge these divisions in the contemporary novel?) |
| Metropolis and Modern city | Nature and the City [35]<br>Metropolis and modernity [36] [53]<br>Abstract Space, French Modernization [45]<br>The rise and decline of the utopian precepts of architectural modernism will be traced<br>The Modernist City<br>The Utopia of the Modernist City<br>The Dystopia of the Modernist City: Metropolis and Warner Bros. Cities of the 1930s<br>The Totality of the Modernist City [32]<br>This course familiarizes students on an advanced level with key concepts and literary texts about the city. It aims to provide insight in and knowledge of the historical connections between modernity and the imagination of urban space [48]<br>Symphonic modernity, complicated modernity, City Symphonies<br>Public Spaces and surveillance (University of Southern California)<br>Modernism in architecture and urban planning [34]<br>Modern Sinophone Cities [31]<br>The city as disciplinary and social control [34] |
| The Country and the City | The City and the Machine: The changing structure of American industry; the relationship of industrialism to urbanization [33]<br>The world has moved from one characterized by rural settlement patterns and provincial lifestyles to one dominated by urbanization, industrialization, immigration, and globalization. [42]<br>The Dark Side of the Small Town, the small town versus the city<br>The Small Town in American Imagination<br>The enduring myth of the small [33] |

Table 2: Codes for Perceived space

**Perceived Space**

| Codes | Code Segments |
|---|---|
| Diversity and Inclusion | The Quintessential American City? [46]<br>Both the city and the cinema grew with the influence of massive waves of immigrants: in the case of the city composing a rich stew of ethnic flavors; in the case of the film industry providing entrepreneurial life blood as well as audience<br>The Immigrant, The Cinema and the City: Course Overview [33]<br>North America: Race Relations<br>North America: Immigration and Assimilation<br>Europe: Changing Demographics and Immigration<br>Ethnic Violence Its Underlying Causes and Manifestations [47]<br>A War on Drugs or a War on the Racialized Working-Class? [49]<br>The cinematic city is an arena of social control and discipline, but also a site of pluralistic diversity, historical memory and social liberation; a society of the spectacle but also a place to make and remake one's self [32]<br>The city provided the technical advances, a vibrant soundstage of streets and crowds, and varied lifestyles to mine for dramatic conflict<br>Many American cities doubled their populations; millions of South and East European immigrants brought their unfamiliar languages, religious institutions and cultural customs to create diversity such as the nation had never before seen {33}<br>Race and Space, Race and Representation<br>What does it mean to be at home in a city versus a foreigner, and at what points do those lines blur? How do cities raise questions of belonging along different scales—from the neighborhood to the national to the transnational or global? [42]<br>Race relations in 1980s New York [34] |
| Technical Cinematic Techniques | Virtual spaces/spaces mediated by technology/imagined or artificial or fictionalized spaces (such as stories within the story, films, books, artworks, or news stories [35]<br>Utopian and Dystopian Cities [34]<br>We will also track the representation of the city into the realm of cyberspace, where it still functions as a site of navigation, perception and self-redefinition<br>The Postmodern city?<br>Cinema and cities both offer utopian built environments of vast perceptual and experiential richness [32]<br>The city of the Future: The urban dystopia; The overdeveloped society (San Jose State University)<br>Dystopias [31]<br>Post apocalyptic Imaginary [50]<br>We will especially attend to the revolutionary impetus in dystopian critiques of urban space and urban utopias posited by the films [51] |

**Perceived Space**

| Codes | Code Segments |
|---|---|
| Gender and Space | Defamiliarization Intersectionality: Race, Class, and Gender [42]<br>Mobilizing Sexual Identity [49]<br>Another strand will explore forces that reshaped sexuality and gender roles in the modern American city and novel<br>A third strand will examine the male literary metropolis [43] |
| Social and Political Issues | The Political Economy of the cities [30]<br>Urban Politics: New Style: The contemporary machine; politics and the media<br>Urban Politics: Old Style: The urban political machine in America; Ethnicity and political participation [33]<br>East Asia (War Ethics) [47]<br>The films will illustrate the development, use, and/or consequences of political power in urbanized areas of the United States, urbanized areas elsewhere (i.e., Paris, and Rio de Janeiro<br>Origins of Contemporary Political Cleavages in the USA [49]<br>The dynamism of metropolis [34] |
| Social and Cultural Issues | Demonstrate an understanding of the key questions of representation and imagination of urban space [48]<br>Social problems in the city<br>Political economy and sociocultural dimensions of cities and urban society<br>Lifestyles in the City: class and the urban process<br>The city as a social backdrop and cultural icon[30]<br>Subcultural Space, Penal Space, Spatial Appropriation [45]<br>Social class and the urban process [33]<br>The Uncomfortable Classroom: Incorporating Feminist Pedagogy and Political Practice into World Regional Geography<br>North Africa/Southwest Asia: Fundamentalist Islam and Women [47]<br>We will investigate the social issues of New York City and the ways in which they are portrayed on film [39] |
| Speed | The City in Motion - Transit and Automobile Space, The American City of the 1970s [45]<br>The Machine in City; The City as Machine [34]<br>Bodies, Machines, and Motion [31] |

Table 3: Codes for Lived Space

| Codes | Code Segments |
|---|---|
| Black Identity | The Harlem Renaissance – Double Consciousness and Urban Black Identity [46]<br>Identities marked by legacies of slavery and colonization [42]<br>American City of the 1970s 2, Blaxploitation and the Ghetto [45] |
| Street | "Violence from Above" and Contentious Politics from Below [49]<br>How many different uses and functions can you find for the city streets (i.e., transportation, site of public protests, etc.)? [42]<br>The street as discursive space [50]<br>As places where the best and worst parts of humanity converge, cities are huge, unwieldy signifiers capable of containing a multitude of meanings—as centers of cultural and economic production, as social experiments, as the sources of moral decay, as breeding grounds for crime, and as sites of multicultural exchange [40] |
| Flaneur | Anti-Architecture, Detective Fiction, flâneur, Labyrinth, Legibility, Madness, Postmodernism, Textuality [46]<br>We will walk alongside narrators and characters as they wander city streets leading through New York, Mumbai, London, Kingston, Brussels,<br>Vancouver, Phnom Penh, Philadelphia, Lagos, and Los Angeles [35]<br>A society of the spectacle but also a place to make and remake one's self (Stanford University)<br>Flaneur, Walking the city<br>The invisible Flaneur [52] |
| Alienation | The Bright City: Coney Island as Urban "Other" [32]<br>Lifestyles in the City: Alienation<br>Urban Anomie [42] |
| Marginalized Groups | How the Other Half Lives – New York and Its Dark Side [46]<br>In what ways do cities open themselves up to some inhabitants and not to others; where are the visible and invisible fault lines that divide, re-route, stall, or prevent access to certain individuals and groups? [42]<br>We will focus on groups of people often marginalized: certain ethnic and racial groups, women, and the economically disadvantaged [47] |

| | **Lived Space** |
|---|---|
| **Codes** | **Code Segments** |
| Contradiction | "New York is a different country," the entrepreneur Henry Ford once quipped. Throughout its history, the 'Big Apple' has been characterized as both an 'independent space' in its own right and the epitome of the 'American Dream.' It is part of this ambiguous rhetoric that the city has come to symbolize 'Americanness' (being the first U.S. capital), but also globalization ('The World's Business Capital'). Perhaps more than any other place in the U.S., NYC lends itself to such contradictory projections [46] <br> The city as represented in literature captures some of the generative and degenerative potential of modern civilization and its unprecedented capacity to foster joy and terror, stimulation and alienation [43] <br> We will also attend to the many ways in which contemporary city novels question the lines between local and global, home and exile [35] <br> South Asia: <br> Tradition and Modernity in a Globalizing World [47] <br> Cities are places to appear and disappear, to emerge and submerge [32] <br> American city, sometimes enhancing, sometimes distorting, sometimes oversimplifying its complexities, exploring its realities, confirming and disconfirming its myths, always adding to the lore of urban life and influencing the popular consciousness of it <br> The celluloid image of urban life is often one of a place of contradictions: of great achievements and failures, hope and despair, community and loneliness, freedom and enslavement, harmony and discord, power and impotence, security and fear [33] |
| Memories and Emotions | The Music of the City – Ragtime, Jazz, and Nightclubbing <br> Brooklyn, Diaspora, Humor, Jewish Community, LA vs. NYC, Art Spiegelman. <br> The Romantic City [46] <br> Sites of memory [36] <br> Shock & Sensation [31] <br> The Syncopated City <br> History and Erasure [32] <br> The Syncopated City [34] <br> Sex in the City <br> Sound and Space: Designing for the Horror and Science Fiction Film <br> City as Musical [50] <br> The city as where the heart is [30] |
| Children | Growing Up in the City: Nature vs nurture in the city [33] <br> Kids are not alright [50] |

| | **Lived Space** |
|---|---|
| **Codes** | **Code Segments** |
| Crime and Terror | Organized Crime, the Urban Jungle, and Noir Aesthetics |
| | Asphalt Jungle, Batman series, CSI, Corruption, Police, Prohibition, Reality Effect [46] |
| | Urban crime and ghost stories, and stories that represent the city as a place in which different species converge [43] |
| | Terrorism and drone warfare |
| | Forms of violence, variously defined [35] |
| | Dark cities |
| | The Bright Dark City [32] |
| | Noir cities [31] |
| | The Violent City: crime, violence and urbanization [33] |
| Humans as Subjects | How exactly do cities get not merely mapped but also emplotted? What kinds of urban spaces and city-dwellers become the privileged focus of modern fiction and poetry? [41] |
| | A new world opens up, and we must find ways to orient ourselves and make sense of it all (MIT) |
| | Modern Woman [31] |
| | It will be about cities themselves, how we see them and navigate them and remember them, and how we might redesign them [38] |

As Table 1 shows, for conceived space, the segments that were coded are mostly related to ideologies, rules and regulations which shape and control space. Codes include: Metropolis and Modernism, which refers to modernism as an ideology that has had a profound influence on city shape and also on social relations in the city [68]. Then we have Climate Change and Immigration and Displacement, which are mainly the results of the rules defined by planners. We include Suburbs, the Farm and the City, Small town, and the Country and the City, and Skyscrapers, which are formed by zoning regulations prepared by technicians and planners. We also categorized Media Representations and Cinematic Techniques as conceived spaces since they are also produced by the dominant power of the society [67].

For perceived space, the segments which were reviewed were those that best represent how people remember the space in their daily lives. Codes include Diversity and Inclusion, which mainly relates to immigration and race relations, and the Future City, for how people envisage the future, how technological advancement helps them to imagine the future, and how they live in technological-made spaces like cyberspace. For Speed, we considered segments that demonstrated how driving a car changed people's perception of their surrounding area. Social and Political issues refers to how political issues changed people's behavior. Social and Cultural issues indicate social representation of the city such as social class. In the Gender category, we considered codes that deal with different experiences of people based on sex.

In lived space, segments that were considered were those that demonstrate how individuals shape their own space with their own culture, character, et cetera. Codes include Black identity which refers to how Black people define their presence in urban areas and Streets as a place which people can manipulate and use for protests and social gathering, among other activities. The other code is Flaneur for those segments which are about a person who develops their character in relation to their experience in the city [69], and Alienation for the codes which are related to the sense of isolation in the modern city [70]. Contradiction is considered for the segments that illustrate the differences and conflicts in the modern city. Marginalized Group refers to people who have been invisible in urban decision making process such as some ethnic, race groups, women, etc. We also coded Children as a specific code because of their importance in urban areas and the efforts—such as Growing up in cities [71]—that have been initiated to create more child-friendly spaces. Crime and Terror refers to unsafe and menacing urban spaces. Memories and Emotions touch on any sort of emotional interaction or social bonds that people develop with the city. And finally, Humans as Subjects reflects how humans interconnect with space.

In reviewing the course assignment, the content was coded into five categories shown in Figure 2. According to the chart, the main learning aim of these syllabi is improving social and cultural knowledge that is more related to the "lived space." Even in courses that are taught in the art or film studies department, the film skills which students learn with film and the city courses are about cinematic techniques that can portray people's emotions, experiences, and daily lives during different periods of time.

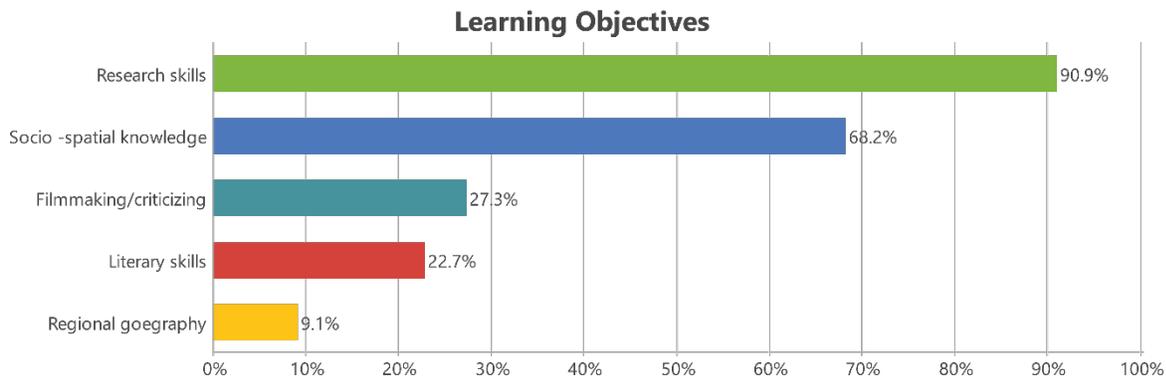

Figure 2: Learning objectives of the twenty-five course syllabi, produced by MAXQDA 2020

Reviewing the course content and learning objectives demonstrated that the main focus of the courses is what we categorized as lived experiences. As mentioned, lived experience refers to the free spaces in which people can create memories, experience different feelings, or roam around. The role of the user in shaping lived experience is very critical. For the assignments, as shown in the Figure 3, the majority consist of a research paper followed by a field trip. The research paper assignment is mostly around analyzing a literary text critically by exploring the representation of the city and the spatial features in it. The research assignment also includes comparing and contrasting city representations in different texts or films, or cinematic techniques that create emotion and meaning. The fieldwork assignment is mostly observing an urban space as described in fictions by using mixed social or other digital media or applying different perspectives. Students are also asked to talk to people in the field, understand their point of view and write up their stories, or to describe a scene of city life that the students have experienced or discover the history of the place by reading plaques, inspirations, old newspapers, or talking to people. Another aspect of the fieldwork assignment is to explore political and aesthetic problems in the city by attending in a political event or meeting.

While both the research paper and the fieldwork assignment aim to engage students in discovering "space" in all of its aspects including lived, conceived, and perceived space, the main focus is on the lived space as students are encouraged to go beyond the physical features of the space and explore how people experience the city.

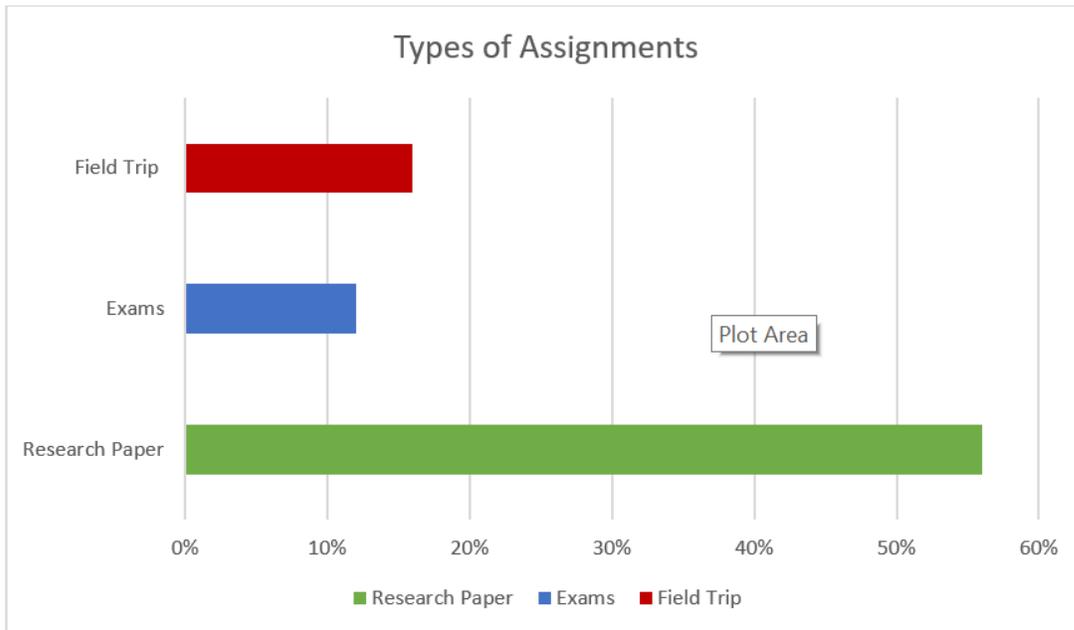

Figure 3: Types of the assignments of the twenty-five course syllabi, produced by MAXQDA 2020

With this in mind, we reviewed the apps to explore the extent to which they considered lived experiences among other spaces in their content and how they approach users in their design. We created our Spatial User Centered Design framework (shown in Table 4) based on codes we defined for the syllabi. The codes incorporate "spatial" information that users need to elevate their spatial experiences. This framework can be used for designing digital platforms related to literary and film geography.

Table 4: Spatial User Centered Design 2(SUCD) Framework

| Types of space | Implications |
| --- | --- |
| Conceived Space | Media as Representational Tool, Skyscrapers, The Country and the City, Displacement and Immigration, Technical Cinematic Techniques, Suburbs, the Farm and the City, Suburbs, Metropolis and Modernism, Climate Change |
| Perceived Space | Diversity and Inclusion, Speed, Social and Cultural Issues, Social and Political Issues, Future City, Gender and City |

| Types of space | Implications |
| --- | --- |
| Conceived Space | Media as Representational Tool, Skyscrapers, The Country and the City, Displacement and Immigration, Technical Cinematic Techniques, Suburbs, the Farm and the City, Suburbs, Metropolis and Modernism, Climate Change |
| Lived Space | Black Identity, Alienation, Marginalized Groups, Contradiction, Flaneur, Street, Children, Crime and Terror, Humans as Subjects, Memories and Emotions |

**4. Analyzing Literary and Film Geography-based Digital Platforms**

We reviewed the digital platforms related to literary geography to explore how these apps are designed and what content they deliver to the users. Our analysis is shown in Table 5.

Table 5: Analyzing map-based apps and interactive maps

| Digital Platforms | Team Expertise | Aim | User Interactivity | Methods of Data Collection |
| --- | --- | --- | --- | --- |
| Mappo/ Mobile app | Initiated by a software developer a navigator and a book lover [72]. Now includes AI, NLP specialists, product designers and business developers [73] | Mapping culture related points of interest (including books, music, podcasts and movies), a point of interest -Presenting uniqueness of every place to pedestrians and drivers [72,73] | Listening to quotes, podcasts, explore and learn more about the place and other contents, share, comment and like, follow recommended and popular routes, create personalized routes, save, and purchase items (novels, films) | Data were collected automatically with a Named Entity Recognition algorithm in machine learning. Fifty thousand points of interest were collected, and people can also add their own points of interest. [72] |

| Digital Platforms | Team Expertise | Aim | User Interactivity | Methods of Data Collection |
| --- | --- | --- | --- | --- |
| JoyceWays/ Mobile app | The project was done by Boston college students in different majors as part of their assignment and their research project. They were responsible for the marketing, designing the layout of the map, photography and organizing the app content [78] (The app is not available on App store anymore) | Joseph Nugent an English professor at Boston College came to this idea to teach *Ulysses* more effectively and also he believed that if students collaborated in making the map, they would produce knowledge and improve the field [78] | Quotations from *Ulysses* with contemporary photos, cartoons, quick facts and images of today (such as bars and pubs which are somehow associated with James Joyce and his works). The app also includes video. People can not comment/like/share or add their own information [78] | Data was collected manually by reading *Ulysses*. 100 locations were identified from 6 chapters of *Ulysses* and were located on the map. They chose 6 chapters which had the richest geographical implications |

| Digital Platforms | Team Expertise | Aim | User Interactivity | Methods of Data Collection |
|---|---|---|---|---|
| StoryTourists/ Mobile app | Initiated by humanities, and a game developer, and continued with marketing and graphic design people [75] | Johanna Forsman, one of the founders, wanted to share one of her favorite books, *Goodbye to Berlin*, with her partner and cofounder of the app, Andreas Jansson. Johanna located the spots where the narrative of the book takes place and had Andreas sit down and read those passages of the book while in those places. Andreas liked the idea and they decided to turn it into business [75] | Solve intricate puzzles, trigger sounds and animations and search for clues in immediate surroundings -Listen to audio books -Camera feature which lets users compare the way the story's city looks at present to the way it looked during the story's time period. There is no function within the app to let users add any type of content [75] | This is a gamified app, and for each story there is an exclusive design. The user takes a role, and the narrator also has a role guiding the user through the tour and helping them to solve puzzles and learn about locations. The data are provided by reading the books that they base their tours on and we then write an adaptation script of the story where they essentially fit an entire book into a 1.5 hour format. This is done manually. Each StoryTour has its own script. Most tours include around 10 points of interaction, where different interactions are triggered depending on what is available in the physical surroundings (statue, plaque, et cetera). All interactions are either puzzles, images, animated sequences, or sounds[83] |
| Toronto Film Map/ interactive map | An assistant media archivist and a geographer and media specialist [79] | The idea came from a similar novel-based interactive map at the Toronto Public Library: The team gathered information based on the available sources in the library and media commons | Users can click on the points and a popup window will show information about the address, the film synopsis, director, production year, and if it is available in the media commons. There is no tab that users can provide information on or change the current information. | They gathered information based on the available sources in the library and media commons [79] |

| Digital Platforms | Team Expertise | Aim | User Interactivity | Methods of Data Collection |
| --- | --- | --- | --- | --- |
| Cultural Atlas of Australia/ interactive map (CAA) | The project was initiated by ARC Discovery Grant. It was carried out by Peta Mitchell – a geography professor and one of the main founders at the university of Queensland | Designed to meet the needs of anyone interested in tracing the ways in which Australian places and spaces have been represented in fictional texts [74] | They click on the places and they can read the name of the place and some information about the novel/play or film. There is a tab in the website where users can contribute to the map making. There are also other sources of information on the website for the places or topics. For instance, in the ShowCase tab users can learn more about ecological themes in Australian films and novels | According to Mitchell [80] close reading of texts that were set in Australia and focused on space and place and were either 1) films that had had a mainstream release, 2) plays that had been performed or organized by a major company, and 3) novels that had won major awards (because there are more novels than films or plays). They also focused on texts that had multiple adaptations (for example, novel and film). |
| | | | | They used the AustLit database to help them determine their sample, particularly when it came to novels, but the sample was ultimately a highly selective and non-representative one, focusing on texts that would provide rich readings for them. They also looked at primary and high school curricula to see which texts were on set reading lists, so their selection was also always thinking about what narrative representations of Australian space young Australians were being exposed to in their schooling [80] |

| Digital Platforms | Team Expertise | Aim | User Interactivity | Methods of Data Collection |
| --- | --- | --- | --- | --- |
| Lord of the Rings interactive map | The developer is a huge fan of Lord of the Rings. His background is in chemical engineering | Emil Johansson, a huge fan of *LOTR*, first published the genealogy, and he received hundreds of emails, from teachers which want to use it for education, or hotels emailed him to put it on the wall of hotel. And then the project found its way to the press. Then he decided to put the timeline next to the map and add more features to the map | Users can see the routes, places, forests, et cetera by clicking on them from the left menu; from the top menu there are tabs which users can explore other non-geographical information such as a family tree, statistics, a geolocated time line of the events, et cetera. Users can not add or manipulate the map, navigation is restricted | The data was gathered by developers' readings of the *Lord of the Rings* books |

According to Table 5, It can be understood that the represented information in these platforms is largely focused on fundamental physical geographical concepts such as relative location, distance, scale, et cetera, reading excerpts of novels, exploring the locations of films pinpointed on the maps, or generally what this paper categorizes as "conceived space" in the syllabi. Some of the apps (Mappo and CAA) [73,74], which let users add places or information to the maps, are mostly representing perceived space as it relates to remembering places that people have already visited and might want to recommend to others. Creating personalized routes is based on the information which already is presented on the map. Restricting users from changing the map or creating their own map has made the user centered experience challenging as people are not able to share their own experiences. In designing these apps, it is hard to maintain "geospatial" information while including "literary" and "film" information, as most of these apps or interactive maps have been developed based on developers' passion for books or films [73,75]. For instance, Dedi Zucker, Mappo CEO, said that they started with the name "Books of Maps," but as they moved forward they realized that their scope is much greater than books [72]; and Lundin [75], marketing and community manager of StoryTourist, says that the "idea behind StoryTourist is that the user should be able to step inside their favorite story. Everyone loves a good story." Depending on the users, the research behind the design is different. For instance, for CAA, the additional information about the points of interest is included in PDF and JPEG photos to give users different environmental, social, and cultural layers of geography to meet their target users, who are researchers. For the other apps, when users click on the points of interest a popup window shows up with information about that point. Teams on these digital platforms are mostly literary analysts, software developers, and marketing analysts. Only two interactive maps [74,76] have cartographers and geographers on their production team and that has also influenced their content creation. As shown in Figure 4, users can get clustered information both in terms of film and geolocation on the Cultural Atlas of Australia interactive map while other platforms do not provide these types of information. In CAA [74] users can also zoom in and zoom out or navigate through the map and find more points of interaction. In Mappo, users can navigate the map in relation to their own location. The app also allows them to choose their journey based on their interests. The app also has a voiceover, which makes it more accessible to different groups of people. StoryTourist and JoyceWays are mainly centered on entertaining users by engaging them in different VR or AR games, as their target users are Joyce lovers and game lovers. The target users for StoryTourists are tourists and the app provides a self-guided literary based app for them. The Lord of the Rings project centers on the specific group of users' needs, and has a great amount of classification of Middle Earth characters, timeline of events, et cetera. However, users' interaction is limited to choosing routes and seeing them on the map, and navigation, as shown in Figure 5, does not give users any extra information. Reviewing these apps reveals that, contrary to literary and film geography theories in academia, the information in these apps is mostly concerned with literary or film information. Some of them include excerpts of the book but without any spatial analysis. These apps mostly prioritize creating an immersive experience for users to delve into stories and films, while they are in the physical spaces in which the story or film takes place.

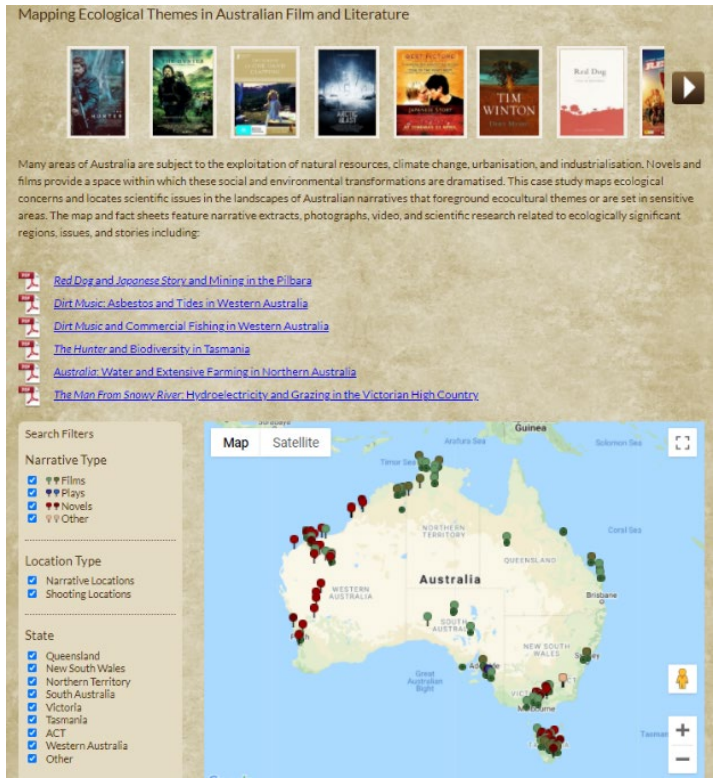

Figure 4: Geolocated types of information offered in Cultural Atlas of Australia, retrieved from: http://www.australian-cultural-atlas.info/CAA/showcase.php?id=EcologicalThemes

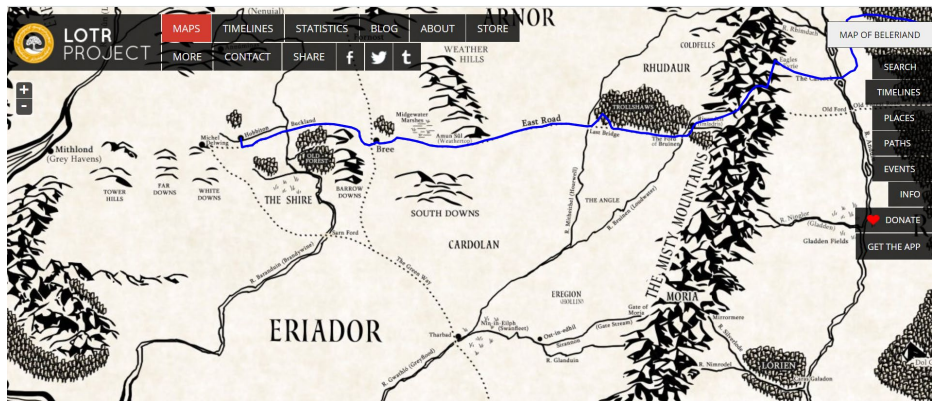

Figure 5: Interaction in the *Lord of the Rings* interactive map is restricted to clicking on the left menu to see the information. Users can not click on the route to get more information. Retrieved from: http://lotrproject.com/map/#zoom=4&lat=-1185.66666&lon=1198.33334&layers=BTTTTTTTT

**5. Analyzing Literary and Film Geography-based Digital Platforms**

Based on the SUCD framework, a model has been designed for a literary/film based app. The focus of the model is on "spatial" knowledge, to help users learn about "space" through a story or a film.

To model the framework, we had to choose a film. To find the most popular movies among possible target users, the first author of the paper lectured on film and literary and film geography to the First-Year Seminar course at UC Davis in Fall 2018, and as part of the lecture asked students to name some films that best represent their lived experiences. *Lady Bird* (2017) was the most frequently mentioned film. The film was watched and the spatial information was categorized according to SUCD framework. Then a prototype of a digital platform was designed in Adobe XD 2020. Figure 6 illustrates the information architecture of the app.

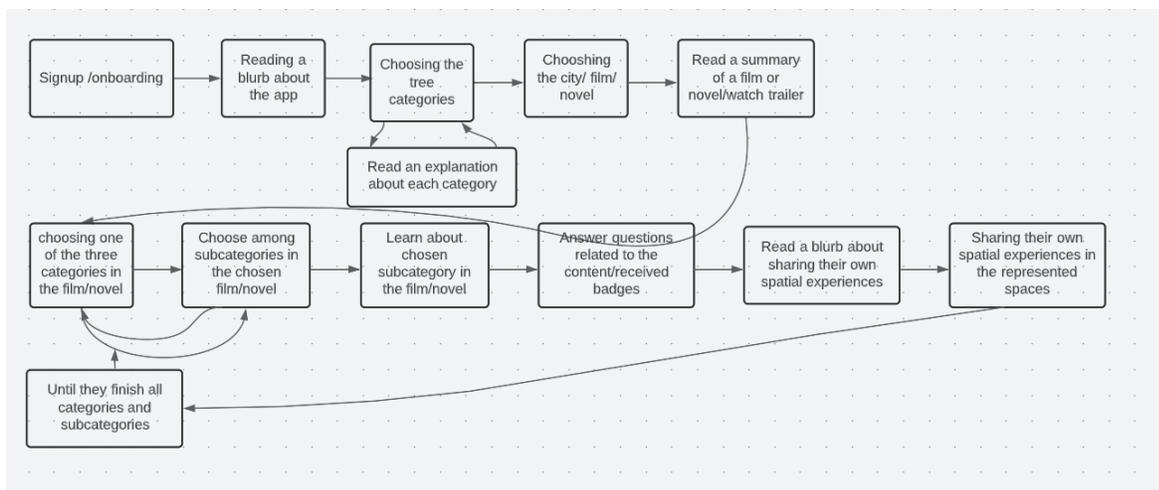

Figure 6: Information architecture of the proposed app

When users sign up, they first read about the idea behind designing the application and then are led to a page explaining the three categories of space. They can choose each category and read the explanation of subcategories. The app will take them to a page where they can select a city, film, or novel. After choosing their preferred medium, they are directed to a page containing a short introduction to the film or novel. On that page, users can explore each of the three categories and subcategories. Each of them represents a specific layer of spatial information.
Figure 7 represents exploring lived space in the film *Lady Bird* in the app.

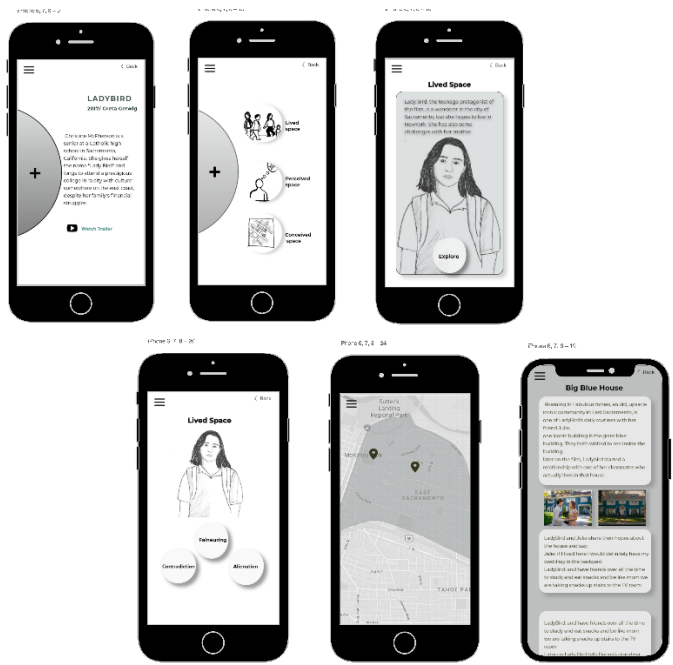

Figure 7: some parts of the user's journey when they choose Flaneur. When they choose Flaneur the following map pops up so they can choose the district or two other locations. By clicking the location for "Blue House" they will be led to Lady Bird and her friend Julie's experiences roaming around the Blue House.

As shown in Figure 7, based on watching the movie, the content was categorized to three subcategories for lived space: Alienation, Contradiction, and Flaneur (shown on table 4), then the content of the app was designed based on these three categories. In Figure 7, the user's journey has been illustrated for the selection of the Flaneur subcategory. To keep users engaged in exploring all different spaces, when they finish a journey, they will deliver a badge (shown in Figure 6). The app also lets them share their own thoughts and experiences. We did our first initial round of expert interviews for the app. According to the interviews, we found that the app can be applied in educational realms, by individuals, and by tourists in the entertainment industry.

## 6. Conclusion

Employment of cartography in map-based applications can bridge the gap between academic literary and film geography and digital practices. The SUCD framework can be applied in geographical learning environments, as well as in tourism and entertainment. It will broaden users' experiences to more than interaction, to learn about "space" that they live in and map their own spatial practices. The proposed SUCD framework empowers users and prepares them for engaging more in making their space. Future research is needed on feasibility of the data collection and analysis of the map using computation methods.

**ACKNOWLEDGMENTS**

We wish to thank Nikki Yang, a senior Sustainable Environmental Design student, for her valuable contributions to the project.